\documentclass[utf8]{frontiersFPHY} % for Physics and Applied Mathematics and Statistics articles

\setcitestyle{square} % for Physics and Applied Mathematics and Statistics articles
\usepackage{url,hyperref,lineno,microtype,subcaption}
\usepackage[onehalfspacing]{setspace}

\newcommand{\grl}{    {\it Geophys. Res. Lett.}}

\newcommand{\jgra}{    {\it J. Geophys. Res. (Space Physics)}}
\newcommand{\jgr}{    {\it J. Geophys. Res.}}

\newcommand{\solphys}{{\it Solar Phys.}}

\newcommand{\ssr}{    {\it Space Sci. Rev.}}
\chardef\us=`\_

\def\keyFont{\fontsize{8}{11}\helveticabold }
\def\firstAuthorLast{Hu {et~al.}} %use et al only if is more than 1 author
\def\Authors{Qiang Hu\,$^{1,*}$, Wen He\,$^{1}$, Lingling Zhao\,$^{1}$ 
and Edward Lu\,$^{2}$}
\def\Address{$^{1}$Center for Space Plasma and Aeronomic Research, and Department of Space Science, The University of Alabama in Huntsville, Huntsville, AL, USA  \\
$^{2}$Department of Electrical Engineering and Computer Science, Massachusetts Institute of Technology, Cambridge, MA, USA}
\def\corrAuthor{Qiang Hu}

\def\corrEmail{qh0001@uah.edu}

\begin{document}
\onecolumn
\firstpage{1}

\title[Evolution of an MC]{Configuration of a Magnetic Cloud from Solar Orbiter and Wind  Spacecraft In-situ Measurements} 

\author[\firstAuthorLast ]{\Authors} %This field will be automatically populated
\address{} %This field will be automatically populated
\correspondance{} %This field will be automatically populated

\extraAuth{}% If there are more than 1 corresponding author, comment this line and uncomment the next one.
%\extraAuth{corresponding Author2 \\ Laboratory X2, Institute X2, Department X2, Organization X2, Street X2, City X2 , State XX2 (only USA, Canada and Australia), Zip Code2, X2 Country X2, email2@uni2.edu}

\maketitle

\begin{abstract}

%%% Leave the Abstract empty if your article does not require one, please see the Summary Table for full details.
\section{}
Coronal mass ejections (CMEs) represent one type of the major eruption from the Sun. Their interplanetary counterparts, the interplanetary CMEs (ICMEs), are the direct manifestations of these structures when they propagate into the heliosphere and encounter one or more observing spacecraft. The ICMEs generally exhibit a set of distinctive signatures from the in-situ spacecraft measurements. A particular subset of ICMEs, the so-called Magnetic Clouds (MCs), is more uniquely defined and has been studied for decades, based on in-situ magnetic field and plasma measurements. By utilizing the latest multiple spacecraft measurements and analysis tools, we report a detailed study of the internal magnetic field configuration of an MC event observed by both the Solar Orbiter (SO) and Wind spacecraft in the solar wind near the Sun-Earth line. Both two-dimensional (2D) and three-dimensional (3D) models are applied to reveal the flux rope configurations of the MC. Various geometrical as well as physical {{{parameters are derived and found to be similar within error estimates for the two methods. These results quantitatively characterize the coherent MC flux rope structure crossed by the two spacecraft   along different paths. }}} The implication for the radial evolution of this MC event is also discussed.

\tiny
 \keyFont{ \section{Keywords:} Magnetic Clouds, Magnetic Flux Ropes, Coronal Mass Ejections, Grad-Shafranov Equation, Force-Free Field, Solar Orbiter, Wind} 
 %All article types: you may provide up to 8 keywords; at least 5 are mandatory.
\end{abstract}

\section{Introduction}\label{sec:intro}
Magnetic clouds (MCs) represent an important type of space plasma structures observed by in-situ spacecraft missions in the solar wind. They have been first identified in the in-situ spacecraft measurements of magnetic field and plasma parameters, and have been studied for decades, based on heliospheric mission datasets \citep{JA093iA07p07217,Lepping1990,1995ISAAB,Lepping1997}. These include the earlier missions such as the Interplanetary Monitoring Platform (IMP), Helios, and Voyager missions. In later times, a number of NASA/ESA flagship missions, including Advanced Composition Explorer (ACE) \citep{1998SSRv...86....1S}, Wind \citep{1995SSRv...71....5A}, Ulysses \citep{1990udbs.book.....L}, and Solar and
TErrestrial RElations Observatory (STEREO) \citep{2008SSRv..136....5K}, have contributed greatly to the study of Solar-Terrestrial physics in general, and to the characterization of MC structures in particular. Generally speaking, the opportunities for one MC structure to be encountered by two or more spacecraft are rare, but when they do occur, it offers a unique opportunity for correlative and combined analysis between multiple spacecraft datasets {{{ (see references below)}}}.

A few such examples include an early study by  \citet{JA086iA08p06673} by using five spacecraft and the series of MC events in May 2007. During 19-23 May 2007, the newly launched twin STEREO spacecraft, Ahead and Behind, i.e., STEREO-A and B, respectively, were separated from Earth by $\sim 6^\circ$ and $\sim3^\circ$, longitudinally, near 1 au. Therefore,  the ACE, Wind, and STEREO spacecraft constellations observed a series of MC events, which enabled a number of correlative studies by using multi-spacecraft measurements \citep{2009SoPh..254..325K,2009JGRAM,2009SoPhM,2010JA015552}. Additionally, a number of studies took advantage of the rare occurrence of radial alignment of mostly two spacecraft separated in heliocentric distances, $r_h$, from the Sun.  For example,  \citet{2007JGRA..112.9101D} studied an MC event and its evolution between the ACE and Ulysses spacecraft when they were separated radially by a distance of $\sim$ 4 au. They found that although the time-series data have evolved significantly between the two spacecraft, a flux rope configuration was still obtained at each spacecraft location and their magnetic field properties were compared. In this research topic collection, Song et al. \citep{song2021} re-examined this event from the perspective of implications for elemental charge states in MCs. Lately, Davies  et al. \citep{Davies_2021} analyzed an MC event detected in-situ by {{{  the Solar Orbiter (SO), Wind, and  Bepi Colombo spacecraft in April 2020}}}, and related to its solar source CME eruption by using the coronagraphic imaging observations from STEREO. We will re-examine this MC event by using the in-situ measurements from both SO and Wind spacecraft. {{We focus on the reconstruction of the magnetic field configurations and characterizations  of the MC flux rope derived from the Wind spacecraft in-situ data.  Quantitative comparison will be made  with the  magnetic  field measurements along the projected SO spacecraft path across the same flux rope structure. }}

One commonly applied quantitative analysis method for MCs based on single-spacecraft in-situ data usually adopts the approach of an optimal fitting to an analytic solution, such as the well-known linear force-free field (LFFF) Lundquist solution \citep{lund}, against  the time series of magnetic field components within a selected interval. These solutions  have limited one-dimensional (1D) spatial dependence, i.e.,  exhibit spatial variation in the radial dimension away from a central axis only. Recently we have improved the optimal fitting approach by extending the Lundquist solution to a quasi-three dimensional (3D) geometry \citep{2021Husolphys,2020HU3DMCGRL}, based on the so-called Freidberg solution \citep{freidberg}. It represents a more general  3D configuration that can account for, to a greater degree, the significant variability in the in-situ measurements of MCs, such as the asymmetric magnetic field profile and sometimes the relatively large radial field component.    An alternative two-dimensional (2D) method has also been applied to in-situ modeling of MCs, by employing the Grad-Shafranov (GS) equation, describing a two and a half dimensional (2-1/2 D) configuration in quasi-static equilibrium \citep{Hau1999,2001GeoRLHu,2002JGRAHu,2004JGRAHu}. This so-called GS reconstruction method is able to derive a 2D cross section of the structure traversed by a single spacecraft, yielding a complete quantitative characterization of the magnetic field configuration composed of nested cylindrical flux surfaces for a flux rope. Such a solution generally conforms to a cylindrical flux rope configuration with an arbitrary 2D cross section. The GS reconstruction method has been applied in a number of multi-spacecraft studies of MCs \citep[see, e.g.,][]{Hu2005,2007JGRA..112.9101D}, including the aforementioned MC events in May 2007 during the earlier stage of the STEREO mission. In addition, it has been widely applied to a variety of space plasma regimes with extended capability and additional improvement \citep{Hu2017GSreview}.

A new era has begun for  solar and heliospheric physics with the launch of the Parker Solar Probe (PSP) \citep{Fox2016} and the Solar Orbiter (SO) \citep{SOmission2020} missions. They will not only yield unprecedented new discoveries of never-before explored territories, but also provide two additional sets of in-situ measurements at different locations in the heliosphere. PSP will plunge closer to the Sun and reach a heliocentric distance below 0.1 au, and SO will provide highly anticipated measurements over a range of heliocentric distances and beyond the ecliptic plane. In this study, we examine one MC event detected during the month of April 2020 by both SO and Wind spacecraft when they were approximately aligned radially from the Sun, but separated by a radial distance of $\sim$ 0.2 au. We present an overview of the event in Section~\ref{sec:overview}. The analysis results by using both the GS reconstruction and the optimal fitting methods are described in Section~\ref{sec:methods}. In Section~\ref{sec:diss}, we discuss the implications for the radial evolution of MCs under the condition of a nearly constant solar wind speed, based on the current event study results. We then summarize the results from this event study in the last section.

\section{Event Overview} \label{sec:overview}%Davies et al. A&A
\begin{figure}
\centering
%\begin{minipage}[b]{0.50\textwidth}
%\includegraphics[width=1.0\textwidth]{dataace110g20.png}
%\end{minipage}%
%\begin{minipage}[b]{0.48\textwidth}
%\includegraphics[width=1.\textwidth,clip]{orbits.jpg}
%\includegraphics[width=1.1\textwidth,clip]{SOMC2020.png}
%\end{minipage}
\includegraphics[width=1.0\textwidth]{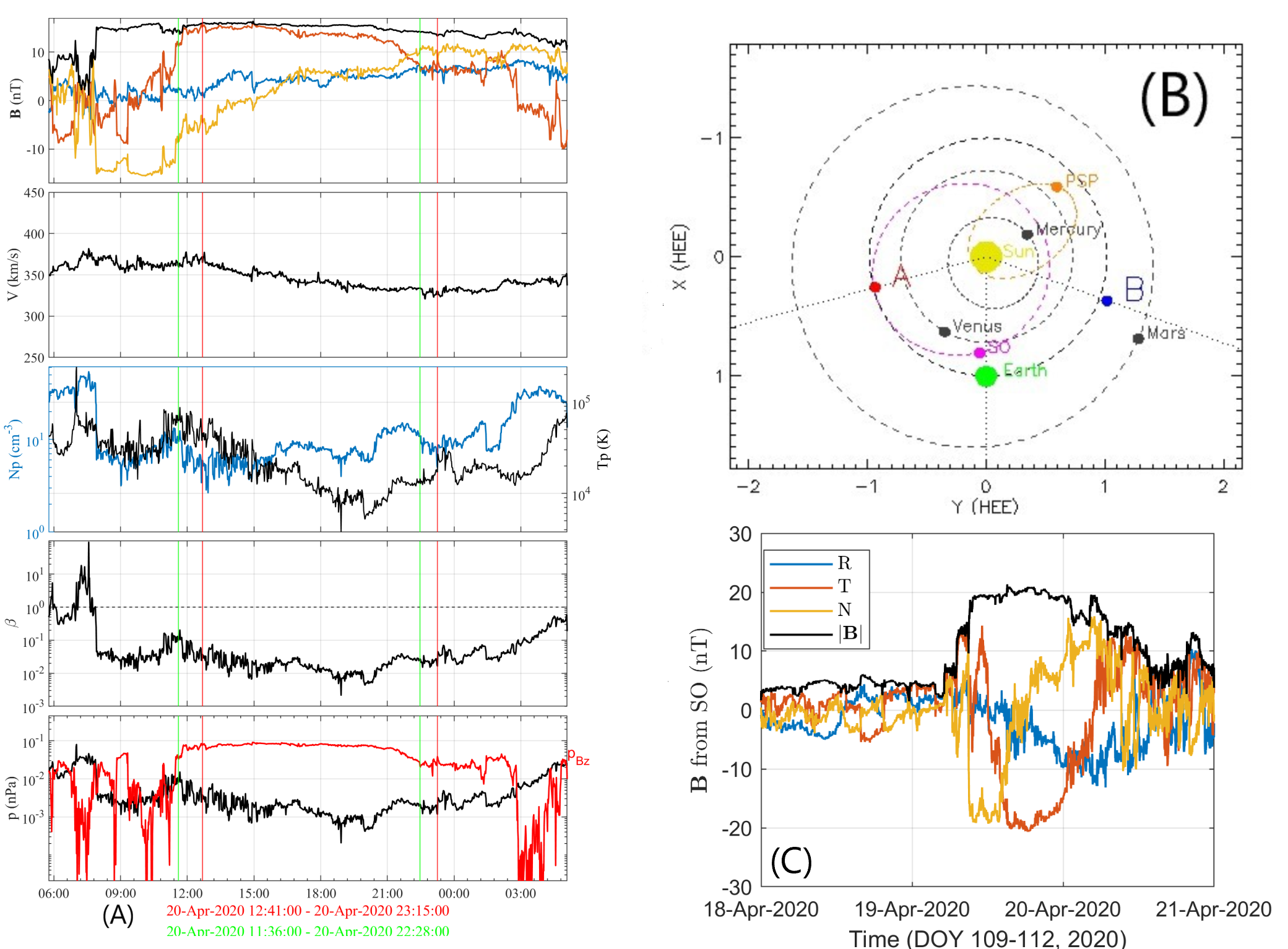}
\caption{(A) Time series data from Wind spacecraft: (from top to bottom panels) the magnetic field components in GSE-X (blue), Y (red), and Z (gold) coordinates and the magnitude (black), the solar wind speed, the proton number density (blue; left axis) and temperature (black; right axis), the proton $\beta$, and the proton plasma pressure and the axial magnetic pressure (red). Two sets of vertical lines mark the intervals for the GS reconstruction (green) and the optimal fitting to the Freidberg solution (red), respectively, and are denoted beneath the last panel. (B) The multiple spacecraft and planets locations around 20 April 2020 in the ecliptic plane (courtesy of the STEREO Science Center). (C) The corresponding SO magnetic field measurements in the RTN coordinates (see legend). }\label{fig1}
\end{figure}

The SO mission observed its first ICME event on 19 April 2020 (day of year, DOY 110) at a heliocentric distance $\sim$ 0.81 au near the Sun-Earth line \citep{Davies_2021,Zhao_2021}.  {{{As summarized in \citet{Davies_2021}, the ICME complex arrived at SO at 05:06 UT, as marked by an interplanetary shock, and followed by a ``magnetic obstacle" 3.88 hours later, which may embody a flux rope structure, and lasted for about 24 hours. }}} The Wind spacecraft subsequently observed the same structures about 1 day later.  Figure~\ref{fig1}A and C show the in situ measurements from the spacecraft Wind and SO (magnetic field only), respectively. Figure~\ref{fig1}B shows the relative locations of a number of objects of interest including SO and Earth (Wind) on the X-Y plane of the Earth Ecliptic (HEE) coordinate system. Relative to Wind,   SO was offset from the Sun-Earth (Wind) line by about 4.02$^\circ$ to the East, while it was North of the ecliptic plane with a latitude of about 1.22$^\circ$ \citep{Davies_2021}.

{{{In Figure~\ref{fig1}A, two intervals are marked for the subsequent analysis of the ICME/MC flux rope structure via the GS reconstruction method (between 11:36 UT and 22:28 UT) and the optimal 3D Freidberg solution fitting approach (between 12:41 UT and 23:15 UT) on 20 April 2020.}}} The in-situ measurements enclosed by the vertical lines indicate clear signatures for an MC: 1) elevated magnetic field magnitude, 2) relatively smooth rotation in field components {{{(i.e., mainly the GSE-Z component varying from negative to positive values)}}}, and 3) depressed proton temperature and $\beta$ value. The corresponding measurements of magnetic field components at SO show similar features with slightly enhanced magnetic field magnitude. The plasma measurements were not available during these earlier time periods of the mission \citep{Davies_2021}. In particular, the rotation in the N component of the magnetic field at SO corresponds well to the rotation in the GSE-Z component at Wind, while the East-West components (along T and the GSE-Y directions) are approximately reversed. {{For a typical cylindrical flux rope configuration crossed by a single spacecraft, the magnetic field component with a uni-polar pattern usually corresponds to the  field component along the axis of the flux rope, while the change in the north-south or east-west component usually indicates the rotation of the transverse field about the axis.   Therefore these signatures, for this particular MC event, hint at a flux rope configuration lying near the ecliptic with the axial direction pointing eastward (positive GSE-Y component, aligned with the thumb of the left hand) with respect to the Sun and with a left-handed chirality (the handedness; GSE-Z component rotating from southward to northward direction, aligned with the other four fingers).}} Given the difference in the magnetic field magnitude and a 1-day time delay consistent with the radial separation distance between SO and Wind \citep{Davies_2021}, it is plausible to consider an evolution between the two spacecraft as well as the spatial variation, assuming that the two spacecraft crossed the same structure along different paths mainly due to  their longitudinal separation.  In what follows, we present our analysis results and discuss the interpretations.

% For Original Research articles, please note that the Material and Methods section can be placed in any of the following ways: before Results, before Discussion or after Discussion.

\section{Methods and Results}\label{sec:methods}
We have developed and applied both 2D and 3D flux rope models to in-situ spacecraft measurements of MCs. The 2D model is based on the Grad-Shafranov (GS) equation and is able to derive a 2D cylindrical configuration with nested flux surfaces of arbitrary cross section shape \citep[see, e.g.,][]{Hu2017GSreview}. The 3D model is based on a more general LFFF formulation, the so-called Freidberg solution \citep{freidberg}, and accounts for a greater deal of variability in the in-situ data through a rigorous $\chi^2$ optimal fitting approach. This approach was recently developed and described in \citep{2020HU3DMCGRL,2021Husolphys}. {{Both methods can yield a set of parameters characterizing the geometrical and physical properties of the structure, including the axial orientation in space, the handedness (i.e., chirality, sign of magnetic helicity), and the axial magnetic flux content (sum of axial flux over a cross-section area), for a flux rope configuration. }} We apply both methods to the Wind spacecraft data of the MC intervals marked in Figure~\ref{fig1}A, and cross-check with the corresponding magnetic field measurements along the separate SO spacecraft path across the same structure.

\subsection{Grad-Shafranov Reconstruction Results}\label{sec:GS}
\begin{figure}[h!]
\begin{center}
\includegraphics[width=1.\textwidth]{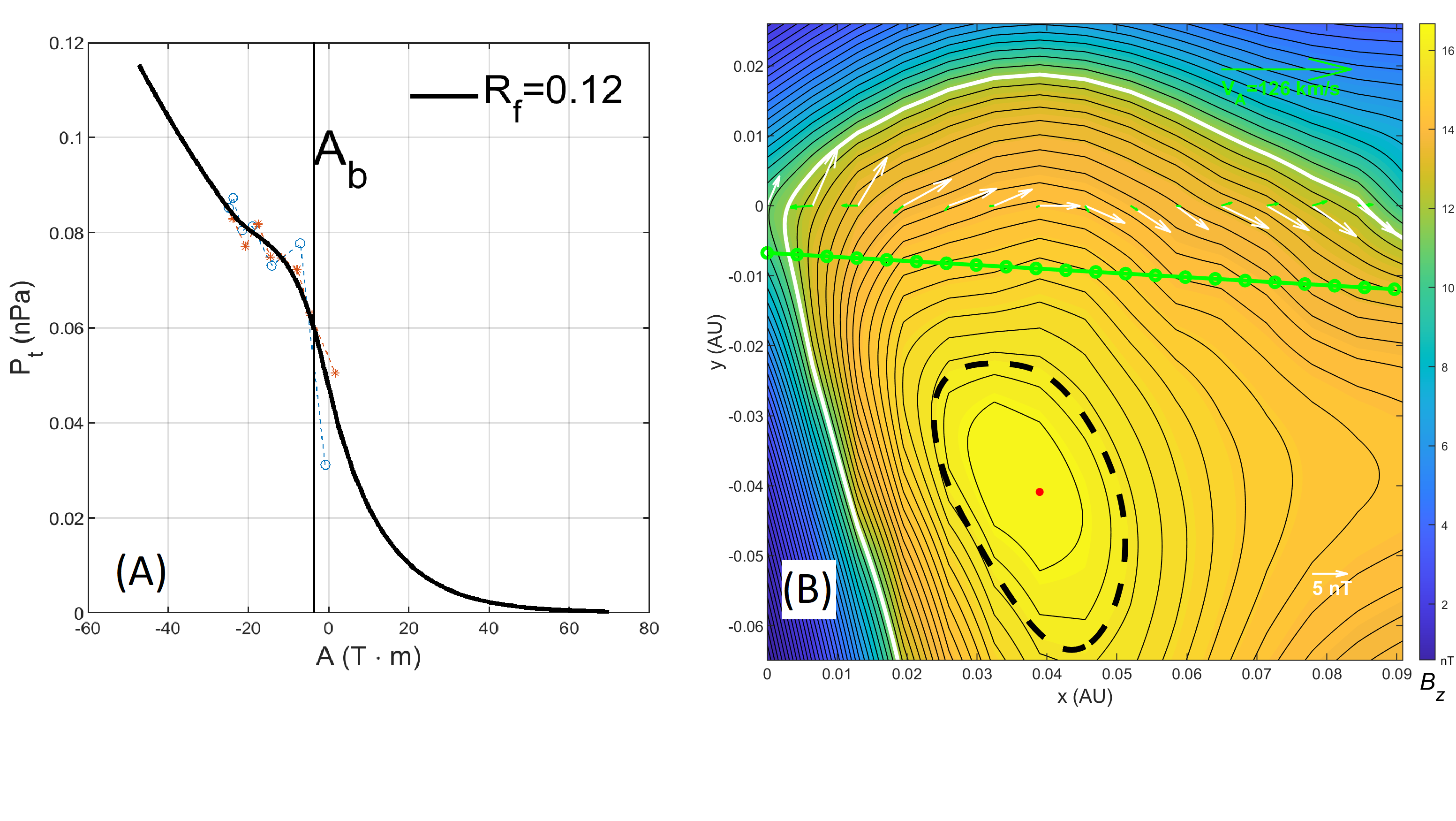}% This is a *.eps file
\end{center}
\caption{(A) The $P_t$ versus $A$ data (symbols) along the Wind spacecraft path, and the fitted $P_t(A)$ curve (thick black curve) with the corresponding fitting residue $R_f$ denoted. The vertical line marks the boundary defined by $A=A_b$, (B) The cross section map from the GS reconstruction for the MC interval marked in Figure~\ref{fig1}. The black contours are the iso-surfaces of $A(x,y)$, and the filled color contours indicate the axial field $B_z(A)$ with scales given by the colorbar. The Wind spacecraft path is projected along $y=0$ with white (green) arrows representing the measured transverse magnetic field (remaining flow) vectors. {{A reference vector proportional in magnitude for each set is provided, respectively, with the white reference vector in the lower right of magnitude 5 nT and the green reference vector of the magnitude of the average Alfv\'en speed in the top right.}} The SO spacecraft path is projected onto the same map as the green line with green circles. The thick dashed contour line highlights the outermost closed loop surrounding the center marked by the red dot where $B_z$ reaches the maximum $B_{z0}$. }\label{fig2}
\end{figure}

The GS reconstruction utilizes the spacecraft measurements of magnetic field $\mathbf{B}$ and solar wind velocity $\mathbf{V}$, and additional plasma parameters as initial conditions to solve the scalar GS equation, which governs the 2-1/2 D magnetic field configuration across the cross section plane perpendicular to the $z$ axis with $B_z\ne 0$ and $\partial/\partial z\approx 0$.  The solution to the GS equation is obtained in the form of a 2D magnetic flux function $A(x,y)$, which fully characterizes the three components of the magnetic field especially including the axial field  $B_z(A)$, among other quantities being single-variable functions of $A$. Figure~\ref{fig2}A shows the data points along the Wind spacecraft path across the MC interval, and the functional form for $P_t(A)$=$p+B_z^2/2\mu_0$, the sum of the plasma pressure and the axial magnetic pressure. Each quantity is a single-variable function of $A$ as required by the GS equation. {{{An optimal $z$ axis orientation is found for which the requirement of $P_t(A)$ being single-valued is best satisfied \citep[for details, see,][]{2002JGRAHu}. For this case,  the  $z$ axis orientation is found to be $(\delta,\phi)=(79,96)\pm(4,9)$\footnote{The polar angle $\delta$ is from the ecliptic north, and the azimuthal angle $\phi$ is measured from GSE-X towards GSE-Y axes, all in degrees.} degrees, with uncertainties estimated by error propagation \citep{2004JGRAHu}.  Then these functions, especially the fitted function  $P_t(A)$,}}} are  used to solve the GS equation and obtain a cross section map of the 2D magnetic field structure given in Figure~\ref{fig2}B for this event. It shows a flux rope configuration with distinct nested flux surfaces (iso-surfaces or contours of $A$), on which the field lines are winding along the $z$ dimension and the axial field  $B_z$ remains the same on each surface. {{The left-handedness (negative chirality) is readily seen from this cross section map, by pointing the thumb of the left hand upward in the positive $B_z$ direction, while wrapping the other four fingers around the direction marked by the white arrows along $y=0$. }} The center of the flux rope defined by the location of the maximum $B_z$ value appears to be away from the spacecraft path at $y=0$ in this case.

\begin{figure}[h!]
\begin{center}
\includegraphics[width=.5\textwidth]{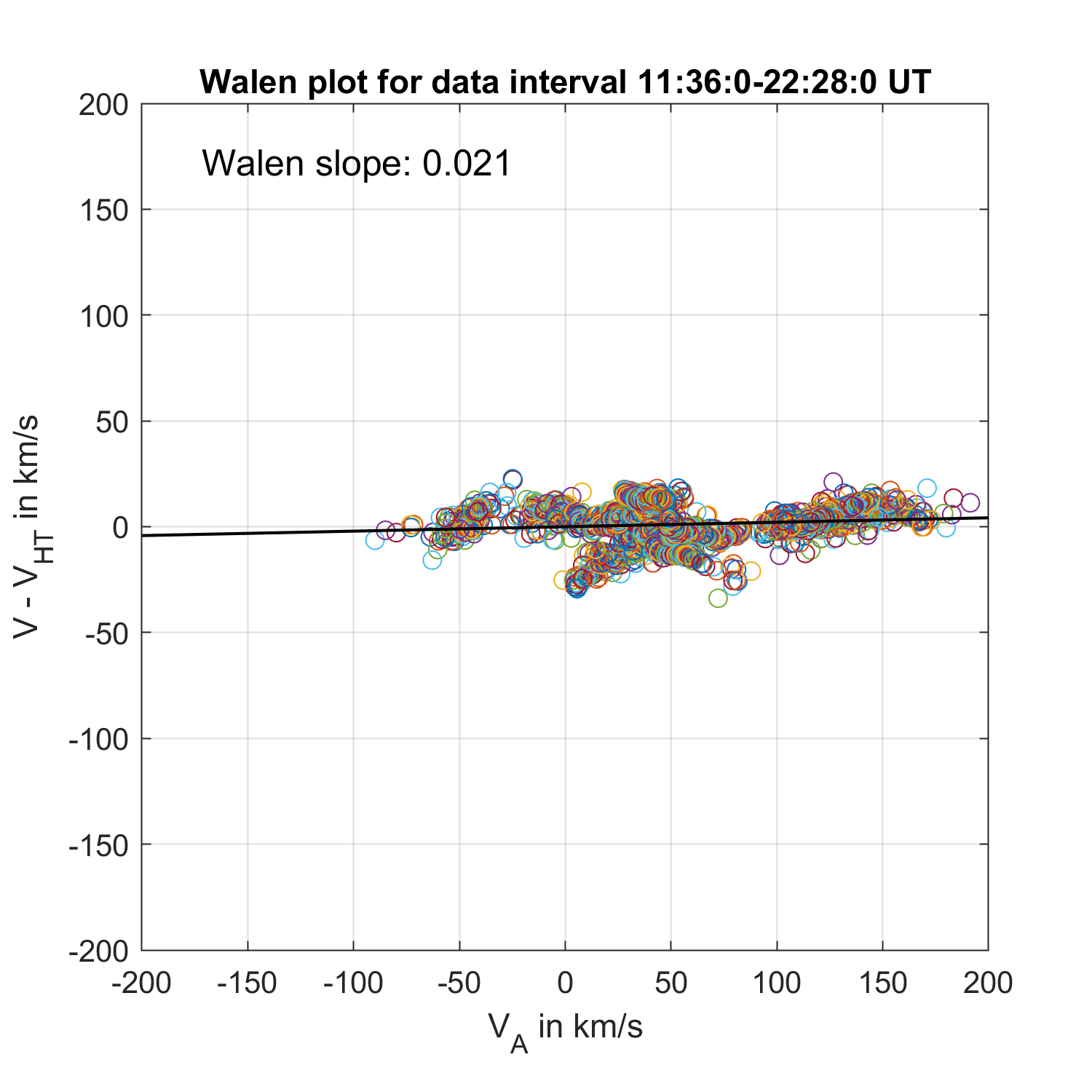}% This is a *.eps file
\end{center}
\caption{The Wal\'en plot for the MC interval at Wind for which the GS reconstruction is applied. The HT analysis yields an HT frame velocity $\mathbf{V}_{HT}=[-346.52, -10.81, 21.10]$ km/s, in the GSE coordinates. The slope of the regression line shown is denoted the Wal\'en slope. }\label{fig5}
\end{figure}

This is  a typical rendering of the GS reconstruction result as viewed down the $z$ axis such that the flux surfaces (contours of $A$) are projected onto the cross-section plane as closed loops surrounding the center for a flux rope configuration. The axial magnetic field usually reaches the maximum at the center and decreases monotonically toward the outer boundary. Along the spacecraft path at $y=0$, the observed transverse magnetic field vectors are tangential to the contours. It is also indicated that {{the remaining flow (green vectors along $y=0$; see also below) as viewed in the frame moving with the flux rope structure is negligible compared with the average Alfv\'en speed (denoted in the top right-hand corner of magnitude 126 km/s).}}  The effect associated with the inertial force in the magnetohydrodynamics (MHD) framework  is assessed via the de Hoffmann-Teller (HT) analysis \citep[see, e.g.,][]{2021Husolphys}.  Figure~\ref{fig5} shows the HT analysis result for this MC interval, in terms of the Wal\'en plot, yielding a slope 0.021 of the regression line. This indicates a negligible ratio between the remaining flow $\mathbf{V}-\mathbf{V}_{HT}$ and the local Alfv\'en velocity.   Thus a quasi-static equilibrium as dictated by the GS equation in the HT frame moving with frame velocity $\mathbf{V}_{HT}$ is approximately satisfied.  For this event, since the SO spacecraft crossed the same structure at a close separation distance but at an earlier time, it is useful to project the SO path onto the cross section map generated by the Wind in-situ measurements, as indicated by the green line with circles in Figure~\ref{fig2}B.  We will further discuss the implications for the radial evolution between SO and Wind in Section~\ref{sec:diss}.

\begin{figure}[h!]
\begin{center}
\includegraphics[width=.8\textwidth]{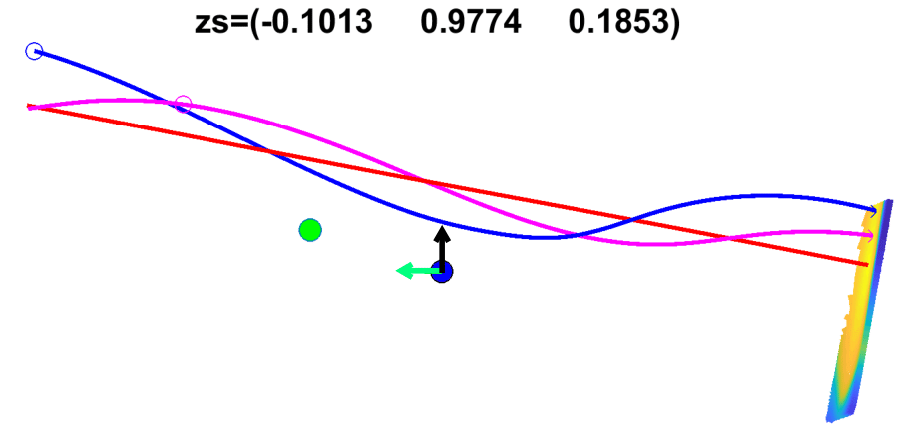}% This is a *.eps file
\end{center}
\caption{The 3D view toward the Sun (along the +GSE-X direction) of the selected field lines (with arbitrary colors) from the GS reconstruction result corresponding to Figure~\ref{fig2}. {{The unit vectors of the GSE-Y and Z coordinates are denoted in green and black arrows, respectively.   The cross section as seen in Figure~\ref{fig2}B is shown to the right where the field lines originate and spiral along the $z$ axis. The red straight field line originates from the center of the flux rope as marked in Figure~\ref{fig2}B. }}The spacecraft locations of Wind and SO are marked by the blue and green dots, respectively. The $z$ axis direction is denoted on top in the GSE coordinates.  }\label{fig3}
\end{figure}

\begin{figure}[h!]
\begin{center}
\includegraphics[width=1.\textwidth]{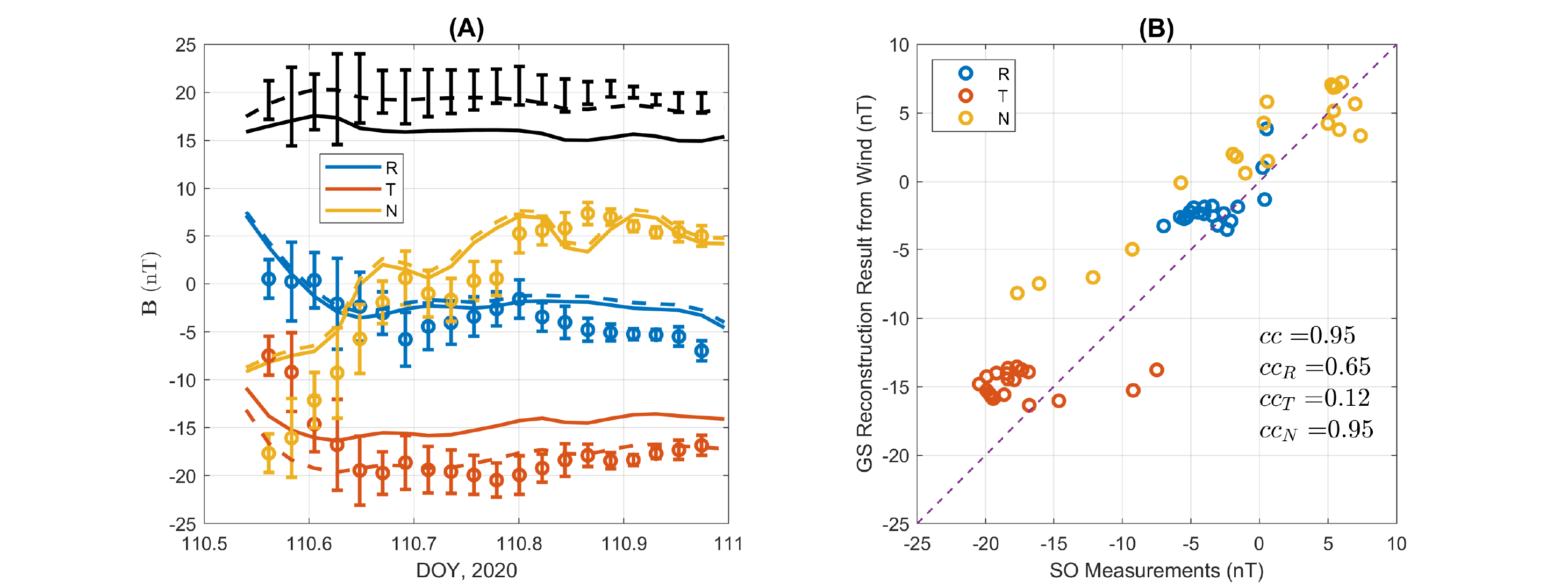}% This is a *.eps file
\end{center}
\caption{(A) The comparison between the derived magnetic field components (solid curves) based on the GS reconstruction result from Wind in Figure~\ref{fig2}, and the actual measurements (circles and error bars), along the SO spacecraft path. The field magnitude is in black. (B) {{{The corresponding component-wise one-to-one scatter plot with the correlation coefficients between the two sets for all three components, $cc$, and each individual component are denoted.}}} {{The dashed line marks the one-to-one diagonal line.}} The dashed curves in (A) represent an alternative estimate/adjustment based on an argument of the $1/r_h$ dependence of the axial field. }\label{fig4}
\end{figure}

It is also informative to illustrate the magnetic field configuration in the perspective view toward the Sun with both Wind and SO spacecraft locations marked in Figure~\ref{fig3}. This provides a direct 3D view toward the Sun {{(located at the same position as Wind in this view but at a distance 1 au away)}} along the Sun-Earth line. It is seen that the reconstructed flux rope structure based on the Wind in-situ data along its path shows {{selected spiral field lines with arbitrary colors winding around a central axis represented by the red straight field line, along the $z$ axis direction, pointing approximately horizontally to the East with both Wind and SO spacecraft passing beneath the center of the flux rope, and separated mostly in the East-West direction.}} With the 2D reconstruction result from the Wind spacecraft, it enables a direct comparison between the derived magnetic field components along the SO spacecraft path, as shown in Figure~\ref{fig2}B, and the actual measured ones returned by the spacecraft. Figure~\ref{fig4} shows such comparison of the three magnetic field components in the SO centered RTN coordinates. Figure~\ref{fig4}A shows the component-wise time series within the MC interval at SO, while Figure~\ref{fig4}B shows the corresponding one-to-one correlation plot, yielding a correlation coefficient $cc=0.95$, {{{for all three components combined. When the correlation coefficients are computed separately for each component, they yield $cc_R=0.65$, $cc_T=0.12$, and $cc_N=0.95$, respectively, as denoted in Figure~\ref{fig4}B. }}}

One main discrepancy is the underestimated magnitude of the $B_T$ component. If one assumes the conservation of axial magnetic flux, it can be established $B_z\propto 1/r_h$ {{(i.e., inversely proportional to the heliocentric distance, $r_h$)}} with the additional supporting evidence of negligible inertial force provided by, e.g., the HT analysis. When this is the case, the dependence of the cross-section area becomes $\propto r_h$, considering largely the angular expansion but little expansion in the radial dimension, for a flux rope configuration with a $z$ axis orientation nearly perpendicular to the radial direction.   The so-called Wal\'en slope as shown in Figure~\ref{fig5}  signifies the relative importance of the inertial force, including the effect of radial expansion, to the Lorentz force in an MHD equilibrium. A small Wal\'en slope magnitude is thus generally a prerequisite condition for the GS reconstruction and the subsequent optimal fitting approach \citep{2021Husolphys}, when they are all based on an approximate magnetohydrostatic equilibrium, sometimes with even stricter additional condition of being force-free. An adjustment based on the argument of the $1/r_h$ dependence of the axial field can be made to the model output at SO location, as shown in Figure~\ref{fig4}A by the dashed curves. This yields a correlation coefficient (between the dashed curves and  circles) $cc'=0.94$, {{{and correspondingly, $cc'_R=0.65$, $cc'_T=0.23$, and $cc'_N=0.95$}}}, although visually they appear to have improved agreement, especially in the $B_T$ component and the magnitude.  We defer additional discussions regarding the radial evolution of MC to Section~\ref{sec:diss}.
%65, T'=.23, N'=.95

\subsection{A Quasi-3D Configuration Based on the Freidberg Solution}\label{sec:3D}
We also apply an optimal fitting approach based on the quasi-3D Freidberg solution to the MC interval denoted in Figure~\ref{fig1}A. For this interval, an HT frame velocity is obtained $\mathbf{V}_{HT}=[-340.95, -4.16, 22.24] $ km/s, in the GSE coordinates, with the corresponding Wal\'en slope -0.0262. The average proton $\beta$ is 0.023. The three magnetic field components of the Freidberg solution in a local cylindrical coordinates $(r,\theta,z)$ are given as \citep{freidberg}, each with dependence on all three dimensions,
\begin{eqnarray}
\frac{B_z(\mathbf{r})}{B_{z0}} & = & J_0(\mu r)+CJ_1(\alpha r)\cos(\theta+kz), \label{eq:B}\\
\frac{B_\theta(\mathbf{r})}{B_{z0}} & = & J_1(\mu r)-\frac{C}{\alpha}\left[\mu J'_1(\alpha r)+\frac{k}{\alpha r}J_1(\alpha r)\right]\cos(\theta+kz), \\
\frac{B_r(\mathbf{r})}{B_{z0}} & = & -\frac{C}{\alpha}\left[k J'_1(\alpha r)+\frac{\mu}{\alpha r}J_1(\alpha r)\right]\sin(\theta+kz). \label{eq:B4}
\end{eqnarray}
Here the solution involves the Bessel's functions of the first kind, $J_0$ and $J_1$. A set of free parameters includes mainly $C$, $\mu$ (the force-free constant, {{sign of $\mu$ representing chirality}}), and $k$, and additional geometrical parameters accounting for the arbitrary orientation and location of the solution domain relative to the spacecraft path. The parameter $B_{z0}$ is pre-determined as the {{{maximum absolute value among all measured magnetic field components over the analysis interval}}} and $\alpha=\sqrt{\mu^2-k^2}$.  It is clearly seen that for $C\equiv 0$, the solution reduces to the 1D Lundquist solution with only $r$ dependence. 

\begin{figure}[h!]
\begin{center}
\includegraphics[width=.5\textwidth]{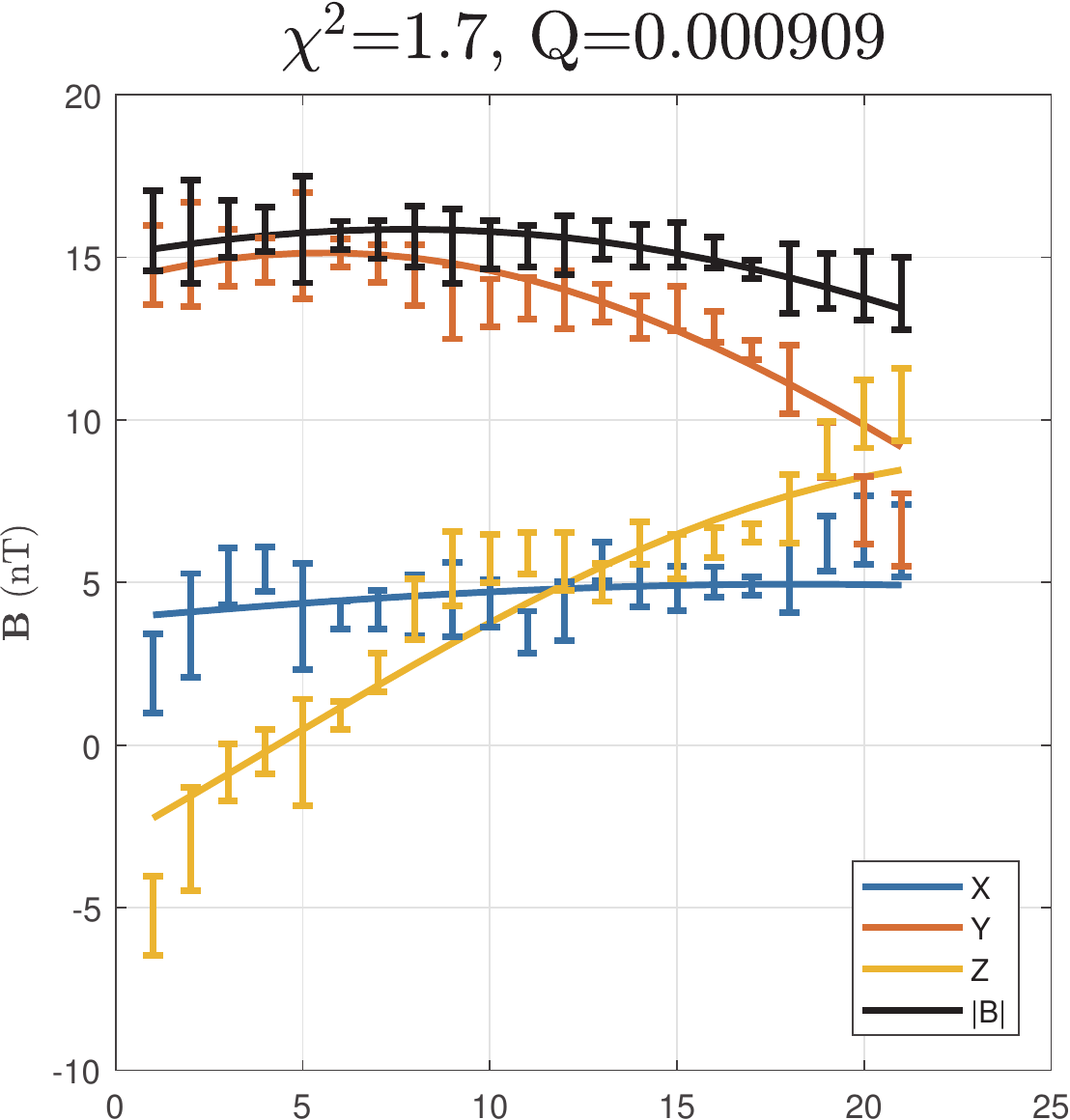}% This is a *.eps file
\end{center}
\caption{The optimal fitting result to the Freidberg solution for the MC interval marked in Figure~\ref{fig1}A. The Wind spacecraft measurements of the magnetic field with uncertainty estimates are shown as error bars, while the corresponding analytic solution is given by solid curves (see legend). {{The horizontal axis is the integral index of the data points. }}}\label{fig6}
\end{figure}

An optimal fitting approach based on $\chi^2$ minimization with uncertainty estimates derived from in-situ spacecraft measurements was devised and applied to a few MC intervals \citep{2020HU3DMCGRL,2021Husolphys}. The results of minimum reduced $\chi^2\lesssim 1$ were obtained in terms of the evaluation of the deviation between the model output from the Freidberg solution and the corresponding spacecraft measurements of the magnetic field components along a single-spacecraft path across the structure. Detailed descriptions of the fitting procedures and comparison of results with the GS reconstruction output and multiple spacecraft measurements are presented in \citet{2021Husolphys}. 
We apply this newly developed approach to the Wind spacecraft data and obtain an optima fitting result as shown in Figure~\ref{fig6}. The minimum reduced $\chi^2\approx 1.7$ is obtained with associated accumulative probability $Q\approx 0.001$, an indication of the quality of the goodness-of-fit, marginally considered acceptable (for $Q\gtrsim 0.001$) \citep{2002nrca.book.....P}.  {{{In addition, the error estimates on the fitted parameters can be obtained via the standard evaluation of confidence limits applicable to such $\chi^2$ minimization as described in \citep{2002nrca.book.....P}. For example, the $z$ axis orientation is found to be $(\delta,\phi)$=(60,90)$\pm(7,9)$ degrees with 90\% confidence limits.  We present the other parameters in Section~\ref{sec:summ}. }}}

\begin{figure}[h!]
\begin{center}
\includegraphics[width=1.\textwidth]{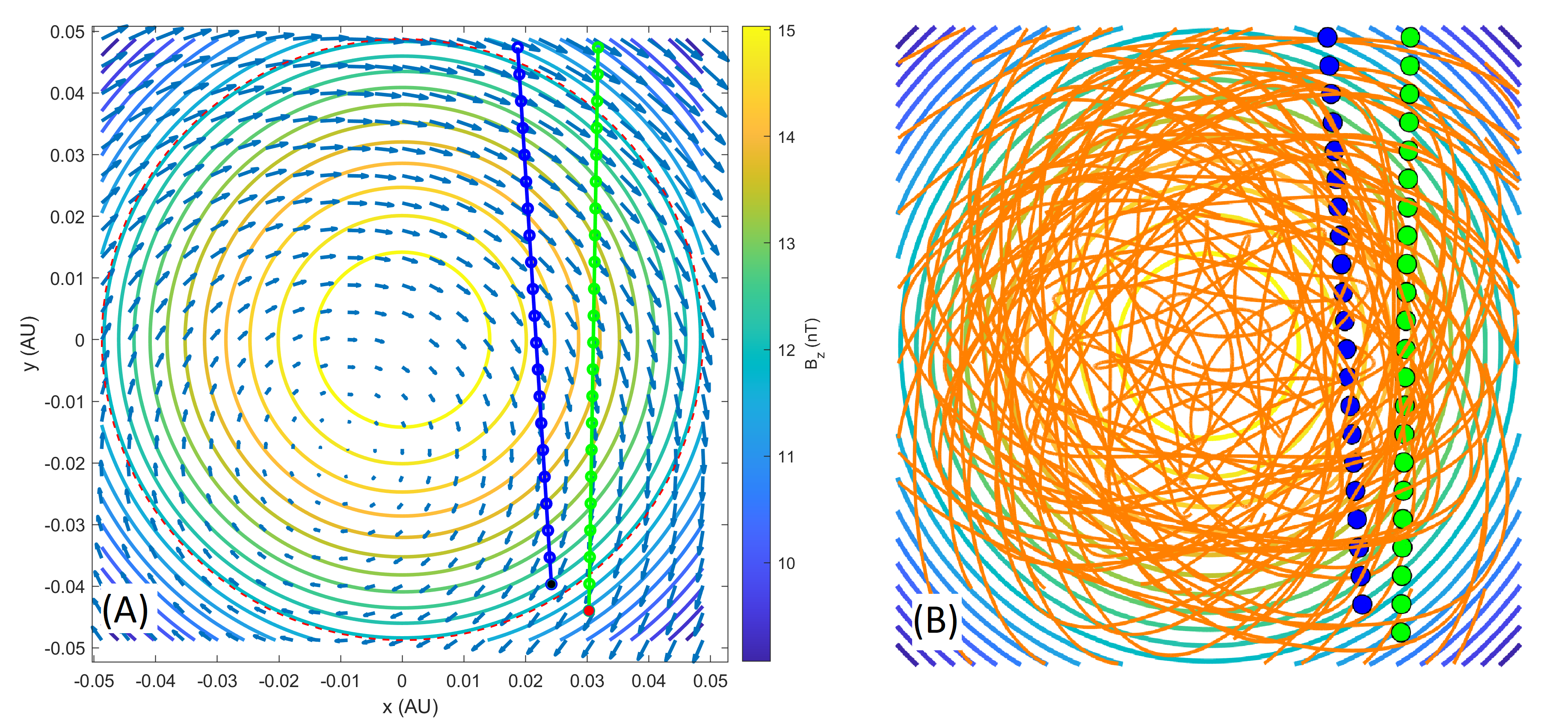}% This is a *.eps file
\end{center}
\caption{(A) One cross section of the optimal Freidberg solution where the colored contours represent $B_z$ distribution and arrows represent the transverse field components. The Wind and SO spacecraft paths are shown by the blue and green lines with dots, respectively. (B) The same view and contour lines for $B_z$ as (A). The orange lines are the field lines originating from the cross section plane, and viewed down the $z$ axis. }\label{fig7}
\end{figure}

\begin{figure}[h!]
\begin{center}
\includegraphics[width=.8\textwidth]{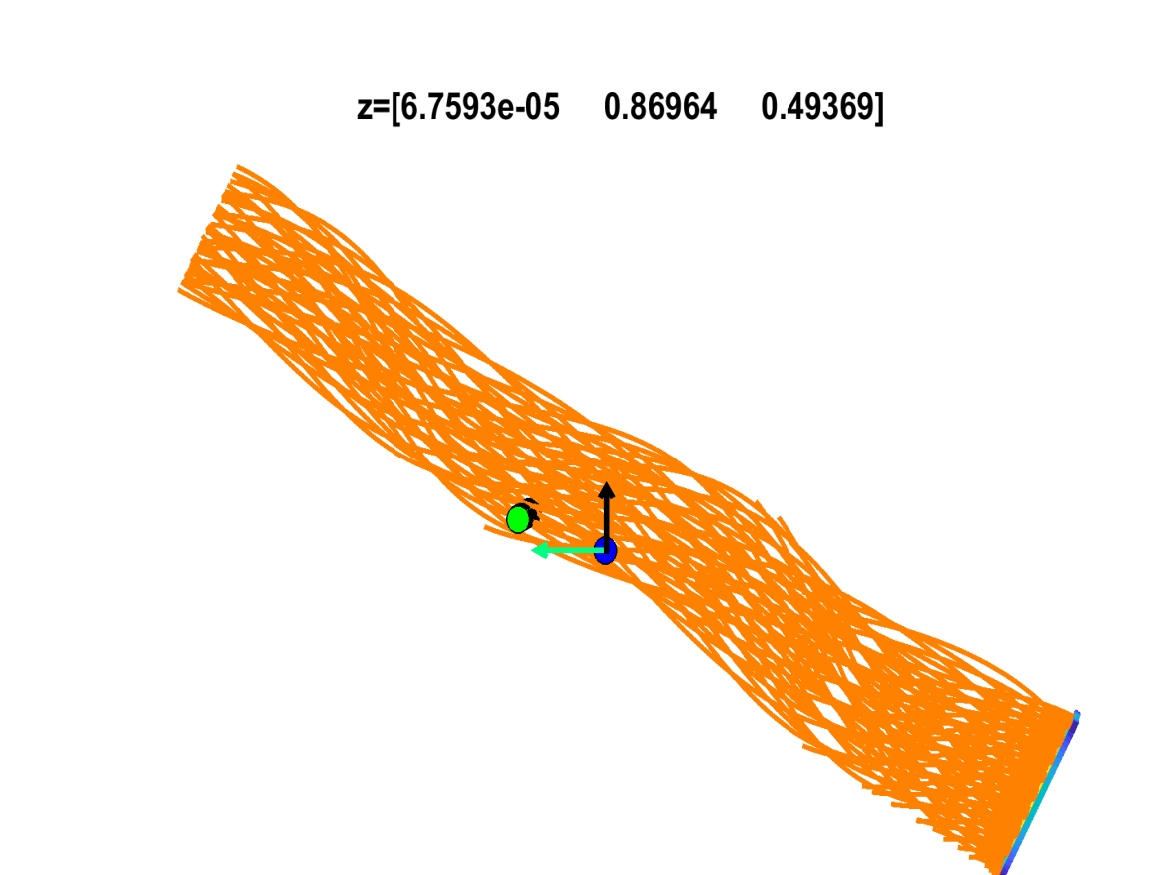}% This is a *.eps file
\end{center}
\caption{The 3D view of the field lines or the flux bundle of the Freidberg solution given in Figure~\ref{fig7}B,  from the same viewpoint and in the same format as Figure~\ref{fig3}. The $z$ axis orientation and the locations of Wind (blue dot) and SO (green dot) spacecraft are also marked.  }\label{fig8}
\end{figure}

\begin{figure}[h!]
\begin{center}
\includegraphics[width=1.\textwidth]{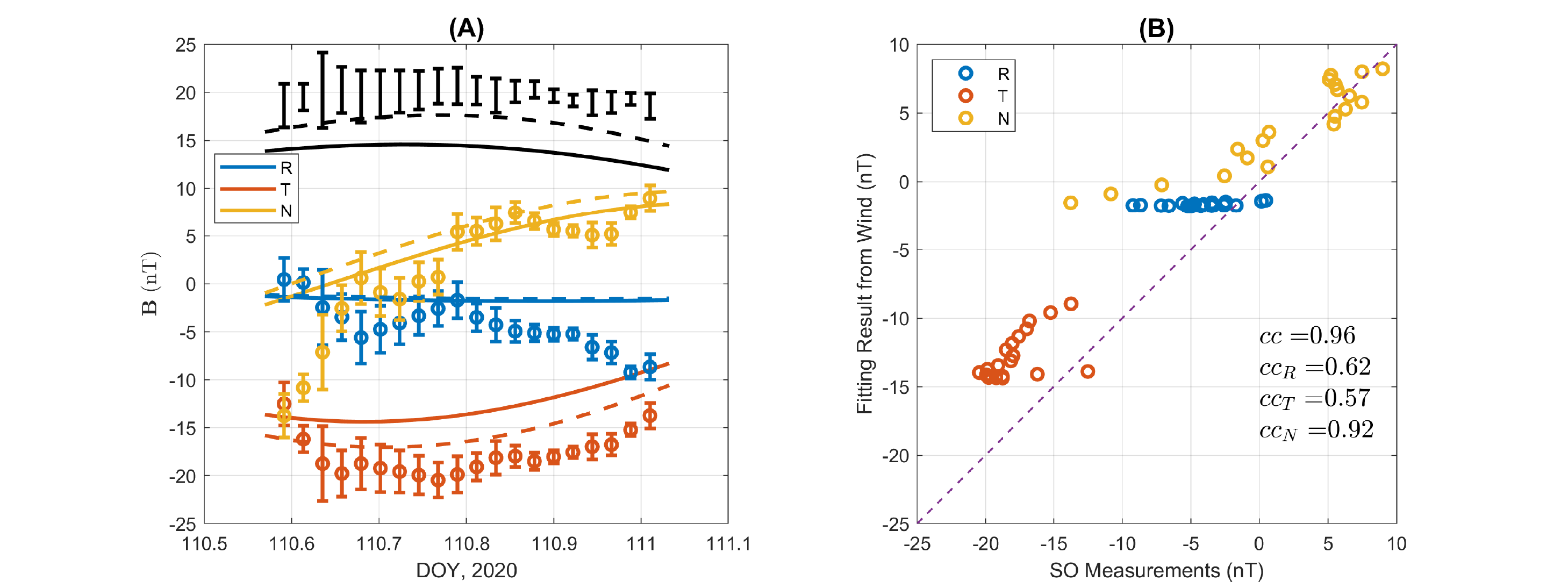}% This is a *.eps file
\end{center}
\caption{The comparison between the derived magnetic field components based on the optimal Freidberg solution fit to Wind spacecraft data, and the actual measurements, along the SO spacecraft path across the solution domain. Format is the same as Figure~\ref{fig4}. }\label{fig9}
\end{figure}

When compared with the GS reconstruction result, the significant distinction of this configuration represented by the Freidberg solution is the 3D nature, not present in any 2D configurations. There no longer exists distinctive 2D flux surfaces, and the field lines exhibit more general 3D features, not lying on discernable individual flux surfaces. Figure~\ref{fig7}A demonstrates one cross section perpendicular to the $z$ axis. The transverse field vectors are not 
tangential to the contours of $B_z$. There is no translation symmetry in the $z$ dimension. To further illustrate this feature, Figure~\ref{fig7}B shows the same view, but with a bundle of field lines drawn in orange color and originating from the bottom plane. No distinctive nested loops (flux surfaces) are seen. As a result, there does not exist a single central field line that is straight along $z$. Figure~\ref{fig8} is the same bundle of field lines viewed from the perspective of the Wind spacecraft toward the Sun. The flux bundle possesses an overall winding along the $z$ dimension, likely related to the topological feature of writhe, giving rise to the 3D feature seen. It also contributes to the individual field line twist, which can be evaluated by the means used for the topological analysis of solar active region magnetic field \citep[e.g.,][]{Liu_2016}. The SO spacecraft appears to cross the flux rope bundle mostly to the East of the Wind spacecraft path, apart from a nominal time delay due to the radial separation. Figure~\ref{fig9} shows the comparison in a format similar to Figure~\ref{fig4}, but for the optimal fitting result of the Freidberg solution to the Wind spacecraft data. The correlation coefficient between the field components from the optimal Freidberg solution and those from the actual measurements along the SO spacecraft path is $cc=0.96$ {{{(additionally $cc_R=0.62$, $cc_T=0.57$, and $cc_N=0.92$). The combined correlation coefficient $cc$ remains the same if adjustments are made as represented by the dashed curves in Figure~\ref{fig9}A, while correspondingly, the correlation between each individual component becomes $cc'_R=0.66$, $cc'_T=0.63$, and $cc'_N=0.92$, based on the argument of solely angular expansion  to be discussed in the next section. }}}
% cc'=.96, R'=.66, T'=.63, N'=.92

%\subsection{Figures}
%Frontiers requires figures to be submitted individually, in the same order as they are referred to in the manuscript. Figures will then be automatically embedded at the bottom of the submitted manuscript. Kindly ensure that each table and figure is mentioned in the text and in numerical order. Figures must be of sufficient resolution for publication \href{http://home.frontiersin.org/about/author-guidelines#ResolutionRequirements}{see here for examples and minimum requirements}. Figures which are not according to the guidelines will cause substantial delay during the production process. Please see \href{http://home.frontiersin.org/about/author-guidelines#GeneralStyleGuidelinesforFigures}{here} for full figure guidelines. Cite figures with subfigures as figure \ref{fig:2}B.
%

%\subsection{Tables}
%Tables should be inserted at the end of the manuscript. Please build your table directly in LaTeX.Tables provided as jpeg/tiff files will not be accepted. Please note that very large tables (covering several pages) cannot be included in the final PDF for reasons of space. These tables will be published as \href{http://home.frontiersin.org/about/author-guidelines#SupplementaryMaterial}{Supplementary Material} on the online article page at the time of acceptance. The author will be notified during the typesetting of the final article if this is the case. 
\section{Discussion} \label{sec:diss}
We lay out, briefly,       a consideration for the radial evolution of the MC, given the difference in the average  magnetic field magnitude between SO  and Wind during the MC interval, which can be partially accounted for by the spatial variations \citep[see, also][]{Davies_2021}. Because the solar wind flow speed at Wind shows little variation, the expansion in the radial direction may be negligible for this event (also justified by the small Wal\'en slope as shown in Figure~\ref{fig5}). Therefore by assuming conservation of axial magnetic flux content and a constant angular extent of the MC flux rope cross section, $\Delta \Theta$, the following relation is assumed to be approximately satisfied,
\begin{equation}
\langle B_z\rangle \Delta r_h\cdot r_h\Delta\Theta\sim\Phi_z\approx Const. \label{eq:Bzr}
\end{equation}
Here the average axial field $\langle B_z\rangle$ is obtained over the cross-section area of the flux rope, which is approximated by the product $\Delta r_h\cdot r_h\Delta\Theta$. The cross-section length scale $\Delta r_h$ is approximately constant if there is little change in the solar wind speed such that any inertial effect including expansion can be omitted (again as judged by the Wal\'en slope). Then, it follows that the average axial field $\langle B_z\rangle$ or approximately $B_{z0}$ changes proportionally with $r_h^{-1}$. This seems to be true for this particular MC event (see Table~\ref{tbl:case1}), and also consistent with    \citet{Davies_2021}. {{Specifically, they found that the radial change of the mean MC field strength follows the dependence $\propto r_h^{-1.12\pm 0.14}$. They also concluded that this MC flux rope was not likely undergoing ``self-similar or cylindrically symmetric expansion".}} For this event, from equation~(\ref{eq:Bzr}) and  Table~\ref{tbl:case1}, it is derived $\langle B_z\rangle\approx 15$ nT at 1 au. It should increase to about 18 nT at SO.  {{{From time-series data, the mean (maximum) total magnetic field strength at SO and Wind are 19 (21) nT and 15 (16) nT, respectively.  It also has to be cautioned that all the reconstructions are based on single-point measurements. }}}  In order to further establish this type of relationship, more event studies are needed.

 {\textbf{This study represents one step forward in the direction of quantifying how realistic MC model outputs are, based on one event study with  available two-spacecraft  in-situ observations.}}  Future work would involve additional measurements and analysis based on remote-sensing observations, which will provide characterizations of solar source region (magnetic) properties of certain MC events to help further assess the fidelity of each model.  {The present implementations represent the best effort we have made in accounting for the variability in the in-situ measurements of MCs and proper error/uncertainty estimates of output parameters. Two models employed are deemed complementary and both are worth applying for individual event studies, as judged by the metrics, mainly, the combined correlation coefficients obtained from this two-spacecraft study with $cc>0.9$. {\textbf{In addition, the correlation coefficients for individual components are better for the Freidberg solution as compared to the GS result. When the radial evolution is considered as assumed by the equation~(\ref{eq:Bzr}), the corresponding  correlation coefficients for both methods slightly improve. }} There  also seems to be a tendency that the Freidberg fitting method is more versatile which yields an acceptable solution when the GS reconstruction method fails \citep[e.g.,][]{2020HU3DMCGRL}. Whether this holds for more number of events has yet to be explored.  }

\section{Summary}\label{sec:summ}
In summary, we have examined one MC event in the solar wind by using the in-situ spacecraft measurements from both the Wind and  SO missions located at heliocentric distances $\sim$ 1 au and $\sim$ 0.8 au, respectively. Two spacecraft were largely aligned along the Sun-Earth line and nearly on the ecliptic plane, but SO was to the East of Wind with a longitudinal separation angle  of $\sim$ 4$^\circ$. The magnetic field measurements from both spacecraft show strong signatures of a magnetic flux rope configuration. In particular, the Wind plasma (not available from SO) and magnetic field measurements confirm the identification of an MC interval, which correlates with the corresponding magnetic field measurements at SO subject to a nominal time delay \citep[see, also,][]{Davies_2021}. We apply both the 2D GS reconstruction method and the optimal quasi-3D Freidberg solution fitting method to the Wind spacecraft measurements and obtain the characterizations of the magnetic field topology at 1 au. A set of parameters  from the analysis is summarized in Table~\ref{tbl:case1}. {{{The error estimates of the parameters for the Freidberg solution are obtained at the 90\% confidence limits, except for $B_{z0}$ and $\Phi_z$. The former is pre-determined and fixed, while the latter is not a free fitting parameter. For the GS result,  an uncertainty range for $B_{z0}$ is also obtained, while the parameters $C$, $\mu$, and $k$ are not applicable ($k=0$ for being 2D). }}} Both methods yield a flux rope configuration with left-handed chirality (``$-$") and their axial directions are oriented mainly along the  West-East direction, with inclination angles relative to the ecliptic plane, about 11$^\circ$ and 30$^\circ$, respectively. The axial magnetic flux content is 1.5-2.1$\times10^{20}$ Mx, and 2.7-2.8$\times10^{20}$ Mx, respectively, {{{as indirectly derived from the model outputs, taking into account the uncertainties}}}. Although the lack of plasma data from SO prohibits the same types of rigorous analysis at SO, we use the available magnetic field measurements at SO to correlate with the corresponding model outputs from the aforementioned quantitative analysis based on the Wind spacecraft data. {{{This becomes feasible for this event study when the two spacecraft were positioned  with an appropriate separation distance. We conclude that both spacecraft crossed the same structure exhibiting a flux rope configuration, as characterized by the set of parameters summarized above. Such an interpretation is supported by the analysis result that the combined correlation coefficients for the GS reconstruction result and the Freidberg solution fitting result are 0.95 and 0.96, respectively.}}}

It is worth noting that as multi-spacecraft measurements become increasingly more available, as partially illustrated in Figure~\ref{fig1}B, new and exciting multi-messenger science will be enabled by using multiple analysis tools. It is highly anticipated that the constellations of current and future missions will usher in new frontiers in heliophysics research.

\section*{Conflict of Interest Statement}
%All financial, commercial or other relationships that might be perceived by the academic community as representing a potential conflict of interest must be disclosed. If no such relationship exists, authors will be asked to confirm the following statement: 

The authors declare that the research was conducted in the absence of any commercial or financial relationships that could be construed as a potential conflict of interest.

\section*{Author Contributions}
QH carried out the analysis and wrote the draft of the manuscript. WH helped with the visualization of the analysis results. LZ obtained the time-series data from SO and participated in the interpretation of the results. EL helped with the analytic verification of the Freidberg solution. 
%The Author Contributions section is mandatory for all articles, including articles by sole authors. If an appropriate statement is not provided on submission, a standard one will be inserted during the production process. The Author Contributions statement must describe the contributions of individual authors referred to by their initials and, in doing so, all authors agree to be accountable for the content of the work. Please see  \href{http://home.frontiersin.org/about/author-guidelines#AuthorandContributors}{here} for full authorship criteria.

\section*{Funding}
Funding is provided, in part, by NASA grants 80NSSC21K0003, 80NSSC19K0276, 80NSSC18K0622, 80NSSC17K0016, and NSF grants AGS-1650854 and AGS-1954503, to The University of Alabama in Huntsville.

\section*{Acknowledgments}
WH and QH acknowledge NSO/NSF DKIST Ambassador program for support.
The authors wish to thank Ms. Constance Hu for proofreading the manuscript. 
We thank the reviewers for useful comments that have helped improve the presentation of this manuscript.

%\section*{Supplemental Data}
% \href{http://home.frontiersin.org/about/author-guidelines#SupplementaryMaterial}{Supplementary Material} should be uploaded separately on submission, if there are Supplementary Figures, please include the caption in the same file as the figure. LaTeX Supplementary Material templates can be found in the Frontiers LaTeX folder.

\section*{Data Availability Statement}
The datasets analyzed for this study can be found in the NASA CDAWeb: {\url{https://cdaweb.gsfc.nasa.gov/index.html/}}.
% Please see the availability of data guidelines for more information, at https://www.frontiersin.org/about/author-guidelines#AvailabilityofData

%\bibliographystyle{frontiersinSCNS_ENG_HUMS} % for Science, Engineering and Humanities and Social Sciences articles, for Humanities and Social Sciences articles please include page numbers in the in-text citations
\bibliographystyle{frontiersinHLTH&FPHY} % for Health, Physics and Mathematics articles

%\bibliography{ref_master3}
% copy .bbl here

%%% Make sure to upload the bib file along with the tex file and PDF
%%% Please see the test.bib file for some examples of references

\section*{Figure captions}

%%% Please be aware that for original research articles we only permit a combined number of 15 figures and tables, one figure with multiple subfigures will count as only one figure.
%%% Use this if adding the figures directly in the mansucript, if so, please remember to also upload the files when submitting your article
%%% There is no need for adding the file termination, as long as you indicate where the file is saved. In the examples below the files (logo1.eps and logos.eps) are in the Frontiers LaTeX folder
%%% If using *.tif files convert them to .jpg or .png
%%%  NB logo1.eps is required in the path in order to correctly compile front page header %%%

%%% If you are submitting a figure with subfigures please combine these into one image file with part labels integrated.
%%% If you don't add the figures in the LaTeX files, please upload them when submitting the article.
%%% Frontiers will add the figures at the end of the provisional pdf automatically
%%% The use of LaTeX coding to draw Diagrams/Figures/Structures should be avoided. They should be external callouts including graphics.

%use minipage to put subfigures side by side, then crop to produce a single figure
% then add (a) and (b), etc by typewriter feature in pdf reader

%\begin{minipage}[b]{0.48\textwidth}
%\includegraphics[height=1.0\textwidth]{FigGS.png}
%\end{minipage}
%\begin{minipage}[b]{0.46\textwidth}
%\includegraphics[height=1.\textwidth,clip]{WIN_ptaace110g20.pdf}
%\end{minipage}

%\begin{figure}[h!]
%\begin{center}
%\includegraphics[width=.45\textwidth]{BmapSO.pdf}% This is a *.eps file
%%\vspace{-3cm}
%\includegraphics[width=.45\textwidth]{WIN_ptaace110g20.jpg}
%\end{center}
%\caption{ (A)  (B) }\label{fig:1}
%\end{figure}

% cc'=.94, R'=.65, T'=.23, N'=.95

% cc'=.96, R'=.66, T'=.63, N'=.92

%\begin{figure}[h!]
%\begin{center}
%\includegraphics[width=15cm]{logos}
%\end{center}
%\caption{This is a figure with sub figures, \textbf{(A)} is one logo, \textbf{(B)} is a different logo.}\label{fig:2}
%\end{figure}

% GS results: Bz0: 16, \Phi_z, 2.0;   16,  1.5; 17, 1.8
{\centering
\begin{table}[tbh]

    \caption{Summary of geometrical and physical  parameters for the MC based on Wind spacecraft measurements.}\label{tbl:case1}
    \begin{tabular}{cccccccc}
        \hline
        %\specialrule{1.5pt}{1pt}{1pt} % Thick line.
        Parameters& $B_{z0}$ (nT) & $C$& $\mu$ & $k$ & $(\delta, \phi)^1$ &  $\Phi_z$ (Mx)& Chirality\\
\hline\hline
        GS result & 16-17 & ...&... &0 &  $(79, 96)$   &  1.5-2.1  & $\mathbf{-}$\\
                      &            &    &     &   &  $\pm(4, 9)$             &   $\times10^{20}$    &(left-handed) \\
\hline
    Freidberg sol. &15 & -0.0047& -0.9848 &-0.9845 & (60, 90)   & 2.7-2.8& $\mathbf{-}$\\
                         &     &$\pm 0.0027$&$\pm$0.0098 &$\pm$0.0098&$\pm (7,9)$  & $\times10^{20}$ &(left-handed) \\
    \hline
    \end{tabular}\\ $^1$\small{The polar angle $\delta$ from the ecliptic north, and the azimuthal angle $\phi$ measured from GSE-X towards GSE-Y axes, all in degrees.}
\end{table}}
%R0=0.04875 au
%<\beta>=0.02 phi_p_2AU =1.297745981459609e+13
\end{document}